\documentclass[fleqn,twoside]{article}
\usepackage{espcrc2}

\usepackage{graphicx}
\usepackage[figuresright]{rotating}

\newcommand{\AmS}{{\protect\the\textfont2
  A\kern-.1667em\lower.5ex\hbox{M}\kern-.125emS}}

\title{MEM  study of true flattening 
 of free energy  and the $\theta$ term\thanks{Poster presented by H. 
Yoneyama}\thanks{This work is supported in part by  Grants-in-Aid for 
 Scientific Research   (C)(2) of the JSPS (No. 15540249) and of the Ministry of Education, Culture,  Sports, Science
 and  Technology (Nos. 13135213 and 13135217).}
 \thanks{SAGA-HE-215,YGHP-04-34}}

\author{Masahiro Imachi\address{Department of Physics, Yamagata 
University, Yamagata 990-8560, Japan},
        Yasuhiko Shinno\address{Graduate School of Science and 
Engineering, Saga University, Saga 840-8502, Japan}
        and
        Hiroshi Yoneyama\address{Department of Physics, Saga 
University, Saga 840-8502, Japan}}
       
\begin{document}

\begin{abstract}
We study the sign problem in lattice field theory with a  $\theta$ 
term, which  reveals as     flattening phenomenon of  the free 
energy density $f(\theta)$.  We report the result of the MEM analysis, where 
such  mock data are used that   `true'  flattening of $f(\theta)$  
occurs .
 This is regarded as a simple model for studying whether the MEM could 
correctly detect non trivial phase structure   
  in $\theta$ space. We discuss how the MEM 
distinguishes fictitious and true flattening.
\vspace{1pc}
\end{abstract} 
\maketitle
\section{INTRODUCTION}
  Lattice field theory with the  $\theta$  term suffers from 
   the sign problem. A conventional technique to circumvent it is to perform the Fourier
   transform of the topological charge distribution
  $P(Q)$ in order to calculate the partition function ${\cal Z}(\theta)$~\cite{rf:Wiese}. 
  The distribution $P(Q)$ is calculated with real positive Boltzmann weight.
  However, this still
  causes  flattening of the free energy, which misleads to a  
fictitious
  signal of a  first order phase transition. To overcome this 
problem requires exponentially 
increasing statistics with lattice volume.\par
     We consider the issue of flattening in terms of  the maximum 
entropy method (MEM)~\cite{rf:Bryan,rf:AHN}.
	  In our previous paper~\cite{rf:ISY} (referred to as (I) hereafter),  we used mock data 
constructed by the Gaussian $P(Q)$ 
	  whose corresponding free energy is analytically known to have   no 
	  singular structure at values  of $\theta$  for $0\leq \theta<\pi$
	  and  tested
   whether the  MEM would be effective  in the case
  with  flattening as well as  without  flattening. 
  In the case without  flattening, the results of the MEM agree 
with the exact results. In the case with  flattening, the MEM gives
  smoother  $f(\theta)$  than that of the  Fourier transform. \par
In the present work, we consider  an opposite case to the above, 
i.e., consider a model  that 
causes  real flattening, which will be  explained later.   This mimics a case 
 where non trivial  phase structure  occurs at finite value of $\theta$ $(\theta < \pi)$. 
 We generate  mock data  based on the  model  and apply the MEM to them.
 Our aim   is to   study  (i) whether 
  flattening would be properly reproduced in this model and (ii)  how 
   true  and fictitious  flattening are distinguished   in the MEM
   analysis.  \par
\section{MEM  AND `TRUE'  FLATTENING}
   The partition function
${\cal Z}(\theta)$ can
  be obtained by Fourier-transforming the topological charge 
distribution
  $P(Q)$: 
\begin{equation}
  {\cal 
Z}(\theta)=\frac{\int[d\bar{z}dz]e^{-S+i\theta\hat{Q}(\bar{z},z)}}
   {\int[d\bar{z}dz]e^{-S}}\equiv\sum_{Q}e^{i\theta Q}P(Q),
    \label{eqn:partitionfunction}
\end{equation}
where $S$ is the action,  and $\hat{Q}(\bar{z},z)$ is the topological charge 
 described by lattice fields $z$ and $\bar{z}$.
 The distribution $P(Q)$ obtained from  MC simulations can be
   decomposed into two parts, a true value and an 
error of $P(Q)$.
When the error at $Q=0$ dominates, the free energy density could develop 
 a flat region in the large $\theta$ region, which   misleads one into 
identifying  flattening  as   a signal of a  first order phase transition at  a finite value of 
$\theta$~\cite{rf:PS,rf:IKY}. \par
   Instead of dealing with Eq.~(\ref{eqn:partitionfunction}), we 
consider
\begin{equation}
   P(Q)=\int_{-\pi}^{\pi}d\theta \frac{e^{-i\theta Q}}{2\pi}{\cal 
Z}(\theta).
    \label{eqn:Pqint}
\end{equation}
The MEM is based on  Bayes' theorem.  In our formulation, in order
to calculate ${\cal Z}(\theta)$, we maximize the  conditional  probability, 
 called the posterior probability
\begin{equation}
   {\rm prob}({\cal Z}(\theta)|P(Q),I,\alpha,m)\propto e^{-\frac{1}{2}
    \chi^{2}+\alpha S},
    \label{eqn:posteriorprobability}
\end{equation}
where the probability is determined by $\chi^{2}$
and the Shannon-Jaynes entropy  $S$  which includes  the default model 
$m(\theta)$
\begin{equation}
   S=\int^\pi_{-\pi} d\theta \biggl[{\cal Z}(\theta)-m(\theta)-
    {\cal Z}(\theta)\log\frac{{\cal Z}(\theta)}{m(\theta)}\biggr].
    \label{eqn:SJentropy}
\end{equation}
 Information $I$ plays the   important role that enforces  ${\cal Z}(\theta)$ to be positive. The parameter $\alpha$ determines the relative 
weights   of   $S$.  
 The best image of ${\cal Z}(\theta)$ for a fixed $\alpha$ is then 
given by 
\begin{equation}
   \frac{\delta}{\delta{\cal Z}(\theta)}(-\frac{1}{2}\chi^2+\alpha S)
    \Bigm|_{{\cal Z}={{\cal Z}^{(\alpha)}}}=0. 
\label{eqn:maximumcondition}
\end{equation}
This  is  followed by the successive procedures in order to obtain 
$\alpha$-independent final image
  ${\hat {\cal Z}}(\theta)$~\cite{rf:Bryan,rf:AHN}:
(i)  Averaging ${\cal Z}^{(\alpha)}$:
\begin{equation}
   {\hat {\cal Z}}(\theta)=\int d\alpha~{\rm prob}(\alpha|P(Q),I,m){\cal 
Z}^{(\alpha)}(\theta).
    \label{eqn:averageofZ2Pa}
\end{equation}
and  (ii)  error estimation.\par
\par
In (I), we used the Gaussian $P(Q)$ for the MEM analysis, which is  
 parameterized as
\begin{equation}
   P_{\rm G}(Q)\propto  \exp[-\frac{c}{V}Q^2] ,
    \label{eqn:Pqparametrize}
\end{equation}
  where $c$ is a constant (=7.42) and $V$ is regarded as the lattice  volume. 
In order to address a question  whether   flattening is due to the
statistical error or due to the characteristic of the  data
themselves,
and a question whether these two are  distinguished  in terms of the 
MEM, we consider  a simple model.
Suppose in some lattice theory that calculated $P(Q)$ 
were  slightly deviated from the Gaussian one only at $Q=0$:
\begin{equation}
 P(Q)=P_{\rm G}(Q)+p_0 \times \delta_{Q,0}, 
\label{eqn:Pqtoy}
\end{equation}
where $p_0$ is a constant. 
    The partition function is analytically obtained  by use of the
Poisson sum formula as
\begin{equation}
   {\cal Z}_{\rm ps}(\theta)\propto 
   \sqrt{\frac{\pi 
V}{c}}\sum_{n=-\infty}^{\infty}
    \exp\biggl[-\frac{V}{4c}(\theta-2\pi n)^2\biggr] +p_0 
	.\label{eqn:poissonsum}
\end{equation}
Because the first term in Eq.~(\ref{eqn:poissonsum}) is monotonically
decreasing function of $\theta$ in the region $0\leq \theta \leq \pi$,
the second one gives   a flat distribution
 at large values of $\theta$. 
 We generate  mock data based on  $P(Q)$   in Eq.~(\ref{eqn:Pqtoy})
by adding    Gaussian   noise with the variance   $\delta \times P(Q)$ for each value of $Q$.
\section{RESULTS}
\label{sec:results}\par
\begin{figure}
        \centerline{\includegraphics[width=7cm, height=6
cm]{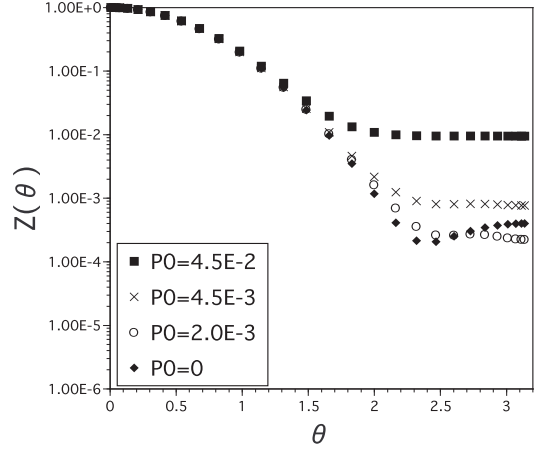}}
\vspace{-8mm}
\caption{Partition function calculated by numerical Fourier transform
for $V=50$. Data for $P(Q)$ are based on Eq.~(\ref{eqn:Pqtoy}) 
with $p_0=4.5\times 10^{-2}$, $4.5\times 10^{-3}$, $2.0\times 
10^{-3}$ and $0$. 
 }
\label{fig:ZV50c0&}
\end{figure}
\begin{figure}
        \centerline{\includegraphics[width=7cm, height=5.5
cm]{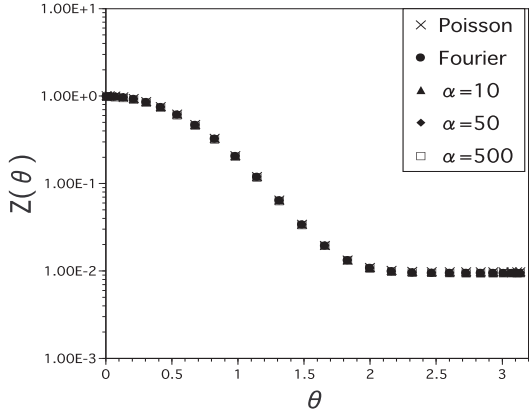}}
\vspace{-8mm}
\caption{${\cal Z}^{(\alpha)}(\theta)$
 for various values of $\alpha$.
Here,  $V=50$ and  $p_0=4.5\times 10^{-2}$.
 }
\label{fig:ZV50c4e-2}
\end{figure}
Figure~\ref{fig:ZV50c0&} displays the behavior of ${\cal Z}(\theta)$ 
calculated by
 the numerical Fourier transform of Eq.~(\ref{eqn:Pqtoy})
 for $V=50$ and various values of $p_0$; $p_0=4.5\times 10^{-2}$, 
$4.5\times 10^{-3}$ and $2.0\times 10^{-3}$.  The parameter 
$\delta$ is chosen to be  $1/400$.
  Figure~\ref{fig:ZV50c0&} also displays  ${\cal Z}(\theta)$ for
 $V=50$ and $p_0=0$, which was studied in (I) in 
detail.
 Clear  flattening is observed in all the cases.
 When  noise is added, the partition function for  
Eq.~(\ref{eqn:Pqtoy})
  is given by
\begin{equation}
   {\cal Z}(\theta)= \frac{\sum_{Q} P_{\rm G}(Q)e^{i\theta 
Q}+p_0+
   \sum_{Q} \Delta P(Q)e^{i\theta Q}}{B},\label{eqn:toyerror}
\end{equation}
where $B$ is a normalization constant. 
When the constant term dominates in the large $\theta$ region,
${\cal Z}(\theta) \approx p_0/B$.
For  $V=50$ and $p_0=4.5\times 10^{-2}$, $B$ is
 numerically estimated as 4.45 and thus
 ${\cal Z}(\theta) \approx 1.0\times 10^{-2} $.
For other choice of $p_0$, $4.5\times 10^{-3}$ and $2.0\times 10^{-3}$, 
$p_0/B$ becomes 
$1.0\times 10^{-3}$ and $4.5\times 10^{-4}$, respectively.
These values agree with those  of flattening  observed in  
Fig.~\ref{fig:ZV50c0&},
 and we thus find that  flattening for  these ${\cal  Z}(\theta)$ is   not caused by  the 
error
 in $P(Q)$ but  is  due to the additional  term in  
Eq.~(\ref{eqn:Pqtoy}). 
On the other hand, flattening for the  $p_0=0$ case  in 
Fig.~\ref{fig:ZV50c0&} is due to the errors in $P(Q)$ as 
investigated in (I).\par
Having distinguished two kinds of flattening, we test whether the MEM 
can properly reproduce the true one.
We employ the case for $p_0=4.5\times 10^{-2}$, as an example.
Figure~\ref{fig:ZV50c4e-2} displays ${\cal Z}^{(\alpha)}(\theta)$ 
for various  values of $\alpha$, 
which is calculated by  Eq.~(\ref{eqn:maximumcondition}).
 The default model is chosen to be the constant one,
 $m(\theta)=1.0$.   We find that ${\cal Z}^{(\alpha)}(\theta)$ 
are  independent  of the values of $\alpha$.  They show clear 
flattening, which are in agreement with
 the result of the Fourier transform and also with that of the 
Poisson sum formula,  Eq.~(\ref{eqn:poissonsum}).
The probability 
${\rm prob}({\cal Z}_n|P(Q),I,m)$  shows 
a narrow  peak located at 
$\alpha\approx 15$. 
Thus $ {\hat {\cal Z}}(\theta)$ agrees with ${\cal 
Z}^{(\alpha)}(\theta)$ in Fig.~\ref{fig:ZV50c4e-2}, 
 and the MEM yields  true flattening.
\par
\begin{figure}
        \centerline{\includegraphics[width=7cm, height=5.5
cm]{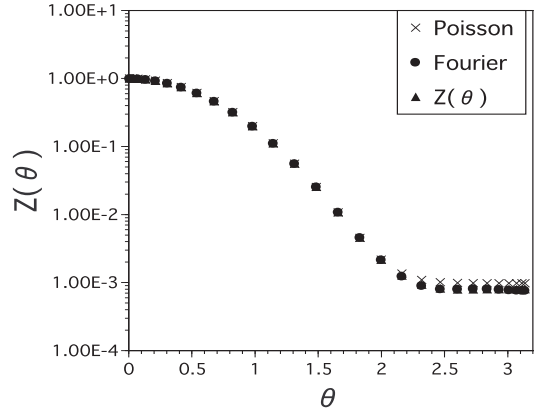}}
\vspace{-8mm}
\caption{$ {\hat {\cal Z}}(\theta)$
for  $V=50$ and $p_0=4.5\times 10^{-3}$.
 }
\label{fig:ZV50c4e-3}
\end{figure}

With   other choice of the value of  $p_0$, we repeat the same 
procedure. 
For $p_0=4.5\times 10^{-3}$, the MEM yields  true flattening as in the 
case $p_0=4.5\times 10^{-2}$. 
This is plotted in  Fig.~\ref{fig:ZV50c4e-3}.
For  $p_0=1.0\times 10^{-3}$, however,  this is not the case.
 We find that the critical value of $p_0$ above which 
flattening is properly reproduced is 
$p_0 \approx 2.0\times 10^{-3}$.  This critical value of $p_0$
 is associated with the magnitude of the error in $P(Q)$. 
\par
In summary, we have applied the MEM to mock data with true flattening and found that 
flattening  is  well reproduced. This is in contrast to the case of fictitious flattening 
studied in (I), where the MEM could predict no fictitious flattening~\cite{rf:ISY}.
 Applicability of the MEM  associated with 
the magnitude of the errors in $P(Q)$ will be reported elsewhere.

\end{document}